# Contact Cluster Modeling of Allosteric Communication in PDZ Domains

Emanuel Dorbath, Fabian Rudolf, Adnan Gulzar, and Gerhard Stock*

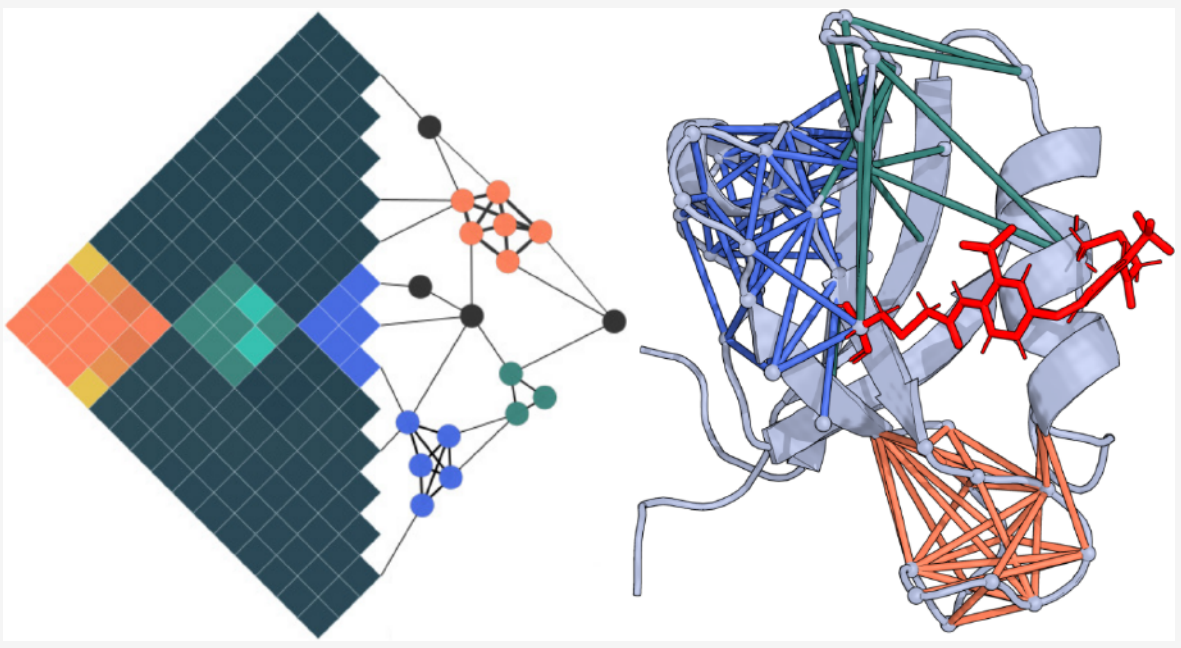

**ABSTRACT:** Allostery, the intriguing phenomenon of long-range communication between distant sites in proteins, plays a central role in biomolecular regulation and signal transduction. While it is commonly attributed to conformational rearrangements, the underlying dynamical mechanisms remain poorly understood. The contact cluster model of allostery [*J. Chem. Theory Comput.* **2024**, 20, 10731−10739] identifies localized groups of highly correlated contacts that mediate interactions between secondary structure elements. This framework proposes that allostery proceeds through a multistep process involving cooperative contact changes within clusters and communication between distant clusters, transmitted through rigid secondary structures. To demonstrate the validity and generality of the model, this Perspective employs extensive molecular dynamics simulations (∼1 ms total simulation time) of four different photoswitchable PDZ domains and studies how different domains, ligands, and perturbations influence both the contact clusters and their dynamical evolution. These analyses reveal several recurring clusters that represent shared flexible structural modules, such as loops connecting $\beta$-sheets, and show that the characteristic time scales of the nonequilibrium protein response can be directly associated with the motions of individual contact clusters. Thus, the dynamic decomposition of PDZ domains into contact clusters uncovers a modular, dynamics-based architecture that underlies and facilitates long-range allosteric communication.

## ■ INTRODUCTION

Allostery is a fundamental mechanism by which proteins regulate essential cellular processes, such as signal transduction and enzymatic activity, by enabling long-range communication between distant protein sites.[1] A comprehensive understanding of allostery requires not just a static picture of protein structure, but insight into both the protein's structure and its conformational dynamics mediating functional coupling.[2−9] However, the direct observation of allosteric transitions remains difficult, particularly because of the subtle nature of structural changes[10−12] and due to the limited sampling of molecular dynamics (MD) simulations.[13−17] Hence, allosteric communication is most commonly explained through network models,[18,19] which describe protein residues as the nodes of the network, while the edges represent inter-residue couplings. Although these models have significantly advanced our understanding of allostery, they do not provide the real-time evolution of the allosteric transition, which is the ultimate goal here.

PDZ domains have become a minimal model for investigating allosteric communication, because of their well-characterized allosteric properties and compact size.[20−27] They share a conserved secondary structure of two or three $\alpha$-helices and six $\beta$-strands. A prominent example is the third PDZ domain (PDZ3, Figure 1a), which exhibits coupling between its ligand-binding pocket and the $\alpha_3$-helix at the C-terminal end. This was shown in an NMR study by Petit et al.,[21] where removing the $\alpha_3$-helix caused a 21-fold reduction in ligand-binding affinity, providing strong evidence for intradomain allosteric communication. To achieve a time-resolved view of this process, Hamm and co-workers[28] incorporated a photoswitch into the $\alpha_3$-helix, triggering a conformational change in PDZ3 that propagated through the protein and caused a distant response within 200 ns. The group further performed time-resolved infrared experiments on other photoswitchable PDZ domains, revealing complex dynamics encompassing multiple time scales over 7 orders of magnitude.[28−30] However, linking these time scales to specific molecular processes has remained challenging.

The experiments described above were paired with extensive nonequilibrium MD simulations,[15,30,31] which, in principle, offer a fully microscopic view of the allosteric transition. The challenge, however, lies in distilling the vast amount of simulation data into a simple, low-dimensional model that still captures the underlying mechanism.[32,33] To address this,

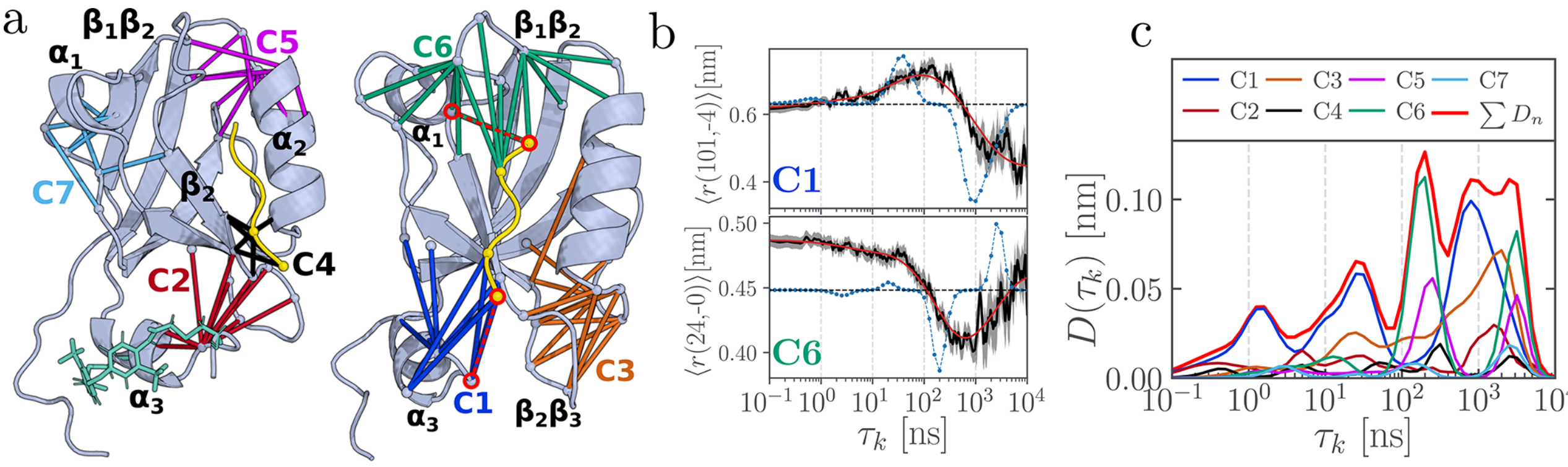

**Figure 1.** Structure and nonequilibrium response of a photoswitchable PDZ3 domain.[28] (a) Illustration including main secondary structural elements, the azobenzene photoswitch (green; shown only on the left), and the ligand (KETWV, yellow) located in the binding pocket between the $\beta_2$-strand and the $\alpha_2$-helix. Colored lines between respective $C_\alpha$-atoms indicate the contact distances associated with clusters C1−C7, as identified by MoSAIC.[45] (b) Time evolution of contact distances $r(101, -4)$ located in cluster C1 and $r(24, 0)$ located in cluster C6; see circled distances in panel (a). (Residues are numbered from 1 to 103 for the protein and from −4 to 0 for the ligand.) MD data are drawn in black, their confidence interval (standard error of the mean) is indicated as the gray area, the time scale spectrum [$a_{kj}(\tau_k)$ in eq 1] in blue, and the resulting fit of the data [eq 1] in red. The increase of the fluctuations in the past decade is due to the reduced number of trajectories for $t \geq 1$ $\mu$s, cf. Table 1. (c) Dynamical content [eq 2] derived from time scale analyses of the contact distances for all clusters combined (bold red) and each individual cluster.

Buchenberg et al.[15] applied principal component analysis to backbone dihedral angles[34] for a PDZ2 domain with a cross-linked photoswitch. Alternatively, Bozovic et al.[30] constructed a Markov state model[35,36] from selected $C_\alpha$-distances, obtained from MD data of a PDZ2 domain bound to a photoswitched ligand. While these models reproduced the experimental time scales, they did not fully resolve the underlying mechanism.

The success of dimensionality reduction and molecular mechanism modeling hinges on selecting initial coordinates (or features) that capture the essential processes.[37−39] This involves two steps: the definition of suitable features and some selection process to discard irrelevant noise. First, guided by comprehensive prior studies,[39−44] we use inter-residue contact distances—encompassing hydrogen bonds, salt bridges, and nonpolar contacts—as our features. We assume that a contact is formed if the distance between the closest non-hydrogen atoms of two non-neighboring residues is shorter than 4.5 Å, see Methods. Because contact distances emphasize short-range interactions, they directly reflect the microscopic mechanism while also capturing long-range structural changes that arise as downstream consequences. (Analogously, in a mechanical machine, the contacts would be the cogwheels of the gearing, and the long-range changes the lever arms performing a function.) Second, in most proteins, only a small subset of these coordinates is involved in a given biomolecular process, making it crucial to discard uncorrelated noise or weakly correlated motions. To achieve this, we apply the Leiden community detection−based feature selection method MoSAIC,[45] which identifies correlated dynamics and partitions proteins into modular subunits—referred to as clusters—based on dynamic connectivity.

To illustrate the explanatory power of these contact clusters, we briefly revisit the above-mentioned experimental study[28] on photoswitchable PDZ3, which revealed a nonequilibrium response on time scales from sub-nanoseconds to tens of microseconds. To interpret these results, Ali et al.[31,46] performed extensive nonequilibrium MD simulations of PDZ3, determined the contact distances $r_j = r(a, b)$ between residues $a$ and $b$, and monitored their time evolution. As representative examples, Figure 1b shows the ensemble average of distances $r(101, -4)$ (between the initially excited $\alpha_3$-helix and the ligand) and $r(24, 0)$ (between the distant $\beta_2$-sheet and the ligand). To characterize the time scales of their evolution, they performed a time scale analysis[47]

$$r_j(t) = \sum_k a_{kj} e^{-t/\tau_k} \tag{1}$$

where the time constants $\tau_k$ are equally distributed on a logarithmic scale (e.g., 10 terms per order of magnitude) and the corresponding amplitudes $a_{kj}$ are fitted to the data, using a maximum-entropy regularization method,[48] see Methods.

Indicated by blue lines in Figure 1b, the analysis reveals that distance $r(101, -4)$ increases and decreases on time scales of 30 and 800 ns, respectively, while $r(24, 0)$ responds not before 200 ns and then again at 3 $\mu$s. The overall response of the system can be quantified via the dynamical content[47]

$$D(\tau_k) = \sqrt{\sum_j |a_{kj}|^2} \tag{2}$$

which sums up the response of all considered features. As shown in Figure 1c, $D(\tau_k)$ reveals peaks at times 1, 30, 200, 800, and 3000 ns, which are found to closely match the experimental results.[28]

To explain these time scales in terms of molecular motion, Ali et al.[46] carried out a MoSAIC analysis[45] on PDZ3 (see Figure S1 and Methods) which block-diagonalizes the contact distance correlation matrix, thus identifying seven localized contact clusters shown in Figure 1a. The contact distances within a given cluster are highly correlated, i.e., they typically exhibit a similar time evolution (Figure S2), although different distances may contribute different time scales. Calculation of the dynamical content $D_n$ for each cluster $n$ (Figure 1c) allowed the assignment of individual peaks to distinct processes: photoswitching of the $\alpha_3$-helix at $t = 0$ initiates its gradual detachment, producing two peaks in cluster C1 at 1 and 30 ns. This is followed by activation of the distant cluster C6 at 200 ns, which in turn induces a back-reaction in C1 at 800 ns, accounting for the realignment of the $\alpha_3$-helix. Finally, at ~3 $\mu$s, structural relaxation occurs in clusters C5, C6 and C7. The contact-cluster framework thus reveals the long-range communication between the initially perturbed cluster C1 and the distant cluster C6, providing a direct mechanistic picture of the allosteric transition in PDZ3.

While the example of PDZ3 provides a clear validation of this strategy, it is important to assess its generality—both in terms of (i) the identification of contact clusters and (ii) their dynamical evolution. If contact clusters are intrinsic features of a protein, we would expect:

- Contact clusters to be identifiable from standard equilibrium MD simulations.
- Similar proteins, such as different PDZ domains, to exhibit similar contact clusters.
- Nonequilibrium studies initiated with different perturbations to be explainable using the same set of contact clusters.

To address these questions, we revisit nonequilibrium studies of various PDZ domains, determine their contact clusters, and examine how different ligands and perturbations influence both the clusters and their dynamical evolution. Specifically, we analyze: a PDZ3 domain bound to a longer ligand (PDZ3L6), a PDZ2 domain with a photoswitch cross-linked over the binding pocket[15,29] (PDZ2S), and a PDZ2 domain with a photoswitchable ligand[30] (PDZ2L). Table 1 summarizes all systems and simulations, while the Supporting Information (SI) provides each protein's sequence and secondary structure, the MoSAIC correlation matrix, the list of contacts included in the clusters, and their nonequilibrium time evolution. We begin by identifying the contact clusters of wild-type PDZ3 (PDZ3WT) from equilibrium MD simulations and comparing them with those obtained from nonequilibrium studies. For this system, we also examine the advantages and limitations of using either nearest heavy-atom contact distances or the corresponding $C_\alpha$-distances as features for defining functionally relevant clusters.

## RESULTS AND DISCUSSION

### Calculating Contact Clusters from Equilibrium Simulations

The contact clusters shown in Figure 1a were derived from nonequilibrium (NEQ) simulations of photoswitchable PDZ3, in which the azobenzene photoswitch was initially switched from a twisted *cis* to a stretched *trans* configuration.[31] In addition to these simulations, Ali et al.[31] performed standard MD simulations of the *cis* and *trans* equilibrium (EQ) states, as well as of the wild-type protein without the photoswitch. Here, we applied MoSAIC clustering to the contact distances of all these trajectories (see Figure S4). As an example, Figure 2a shows the contact clusters identified for PDZ3WT (EQ), which can be directly compared to the clusters of PDZ3 (NEQ) in Figure 1a. In both cases, seven clusters are found, localized in the same regions of the protein. The lists of contact distances are not identical (see Tables S2 and S4), as expected given that the two systems differ in the presence or absence of the azobenzene photoswitch on the $\alpha_3$-helix. Consistently, when comparing contact distances from NEQ and EQ simulations of the same photoswitchable PDZ3 (Table S3), the agreement is even stronger.

**Table 1. Considered PDZ Domains and Equilibrium (EQ) and Nonequilibrium (NEQ) Simulations**

| system | ligand | MD runs |
|---|---|---|
| PDZ3WT[31] | KETWV | EQ: $4 \times 1$ $\mu s$ |
| PDZ3[31,46] | KETWV | NEQ: $90 \times 1$ $\mu s$; $22 \times 10$ $\mu s$ |
| PDZ3L6 | KKETWV | NEQ: $89 \times 1$ $\mu s$; $10 \times 10$ $\mu s$ |
| PDZ2S[15] | – | NEQ: $14 \times 10$ $\mu s$ |
| PDZ2L[30] | RWAKSEAK | NEQ: $80 \times 1$ $\mu s$; |
| | ECEQVSCV | $19 \times 10$ $\mu s$ |

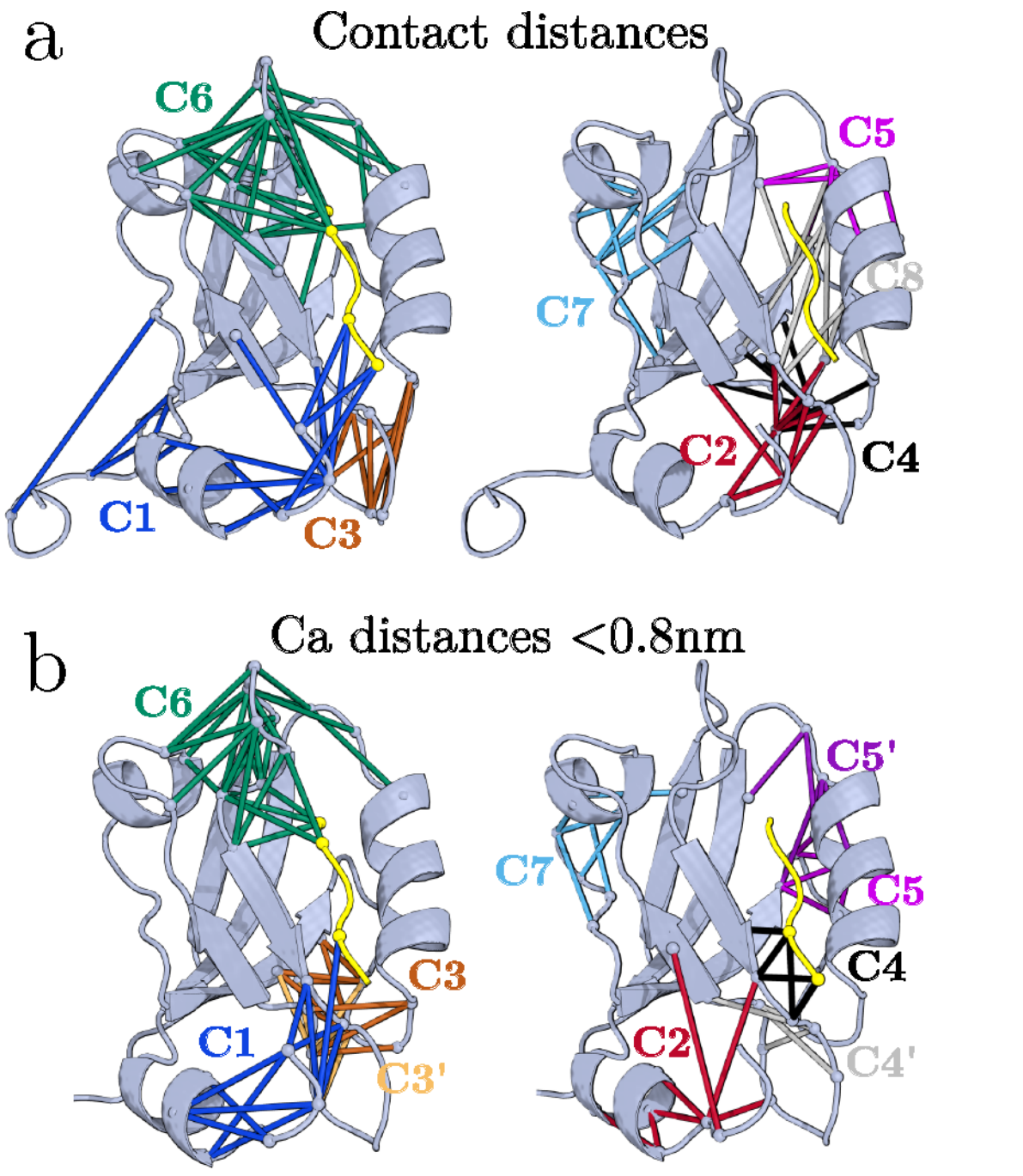

**Figure 2.** Contact clusters C1 to C7 of PDZ3WT obtained from equilibrium MD simulations. Compared are contact clusters obtained from (a) shortest heavy-atom inter-residue distances below 0.45 nm and (b) corresponding $C_\alpha$-distances below 0.8 nm.

We therefore conclude that contact clusters are indeed intrinsic features of a protein, comprising functionally relevant contacts that mediate tertiary interactions between relatively rigid secondary-structure elements. Acting as flexible joints or hinges, these regions play a critical role in enabling structural rearrangements.[49]

### Feature Selection: Contact vs $C_\alpha$-Distances

Several definitions of inter-residue distance can be employed. In our approach, contacts are identified based on the shortest heavy-atom distance between two non-neighboring residues being below 0.45 nm.[42,44] This criterion is computationally demanding, as it requires monitoring all heavy-atom distances for every residue pair throughout an MD trajectory. Using the Python package msmhelper[50] on a standard desktop computer with 10 physical cores, this calculation typically takes about 2 h for a protein of $10^2$ residues and roughly 9 h for a protein of $10^3$ residues, each analyzed over $10^6$ MD frames.

Requiring about an order of magnitude less computation time, we can alternatively monitor the distance between the corresponding $C_\alpha$-atoms, and require this distance to remain below 0.8 nm for at least 10% of the simulation time, see Table S5. To compare these "local" $C_\alpha$-distances with the nearest heavy-atom contact distances, Figure 2b shows the resulting contact clusters for PDZ3WT. In both cases, we obtain seven clusters localized in the same regions of the protein, indicating that local $C_\alpha$-distances provide, on average, a good approximation to the heavy-atom contact distances. Nonetheless, for elucidating the mechanism of a structural change, contact distances are generally preferable, as the formation or breaking of a contact often coincides with local structural stabilization or destabilization. In addition, changes in backbone hydrogen bonds directly affect the amide I band (primarily backbone C═O stretch vibrations), which can facilitate comparisons between MD simulations and infrared spectroscopy data.[12]

Finally, we briefly highlight two alternative approaches for using inter-residue distances as features. First, several studies[51] have discussed the benefits of employing inverse distances, particularly because they offer greater sensitivity at shorter distances. When assessing the dynamical content for this approach, Figure S3 shows that, for PDZ3, the change in dynamical content is minimal. On the other hand, we note that in Markov state modeling it is common to use all $N(N-1)/2$ $C_\alpha$-distances as features.[37,52] For PDZ3WT, we examined this choice in the SI and showed that the resulting main MoSAIC clusters primarily contain distances than span from one or a few anchor points all across over the entire protein (Figure S6). These clusters mostly capture global protein motions, similar to those observed in normal-mode analysis (Figure S7). However, while normal-mode analysis mainly captures nonreactive ground-state fluctuations, our focus is on describing the reaction mechanism driven by contact changes. For this purpose, local $C_\alpha$-distances (i.e., $d \leq 0.8$ nm, see Figure 2b) may be useful, while including all $C_\alpha$-distances tends to be counter-productive.

## PDZ3L6: Modifying the Ligand

We now study how contact clusters and nonequilibrium response of the photoswitchable PDZ3 domain shown in Figure 1 are affected by a small modification of the system. As a test case, we consider the same PDZ3 domain, but with a ligand (KKETWV) that is extended by one additional Lys residue. Performing a MoSAIC correlation analysis on the MD data of this variant (hereafter referred to as PDZ3L6), we again obtain seven contact clusters. Figure 3a reveals that these clusters closely resemble those identified for PDZ3 in Figure 1a (cf. Tables S2 and S6). The only notable (and expected) difference is cluster C1, which exclusively involves contacts with the extra ligand residue Lys(−5): three to the $\alpha_3$-helix, two to the protein core, and one to the ligand residue Thr(−2). Recall that in PDZ3, cluster C1 contains 11 contacts, with several between the $\alpha_3$-helix and various ligand residues (see Table S2). Hence, the addition of a single residue, Lys(−5), completely reshapes the contact network of C1, while leaving the other clusters largely unaffected.

Since cluster C1 is the primary site excited by photoswitching of the $\alpha_3$-helix, it is instructive to examine how the nonequilibrium response of PDZ3L6 differs from that of PDZ3. As an example, Figure 3b depicts the time evolution of the contact distance $r(102, -5)$, which exhibits a single decay on a time scale of ∼80 ns. Accordingly, the dynamical content of PDZ3L6 (Figure 3c) exhibits a first main peak at 80 ns originating from C1, followed by a weaker feature at ∼2 $\mu$s arising from C1 contacts not involving $\alpha_3$ (Figure S9). This is in contrast to the more complex dynamics of C1 in PDZ3 (Figure 1), where the initial detachment of the $\alpha_3$-helix (at 1 and 30 ns) is succeeded by its realignment around 800 ns. Another notable difference of PDZ3L6 is that the response of the distant cluster C6 (distance $r(18, -1)$) occurs already at 100 ns (compared to 200 ns in PDZ3). This acceleration is most likely induced by the intraligand contact $r(-5, -2)$, which provides a shortcut in the mechanical coupling between clusters C1 and C6.[46] At longer times, the dynamical content of PDZ3L6 reflects the combined response of clusters C1, C3, C5, and C6, which closely resembles the behavior observed for PDZ3. In summary, although the details of the short-time dynamics differ from those of PDZ3, the overall long-range coupling between C1 and the distant clusters C5 and C6 evolves in a similar manner.

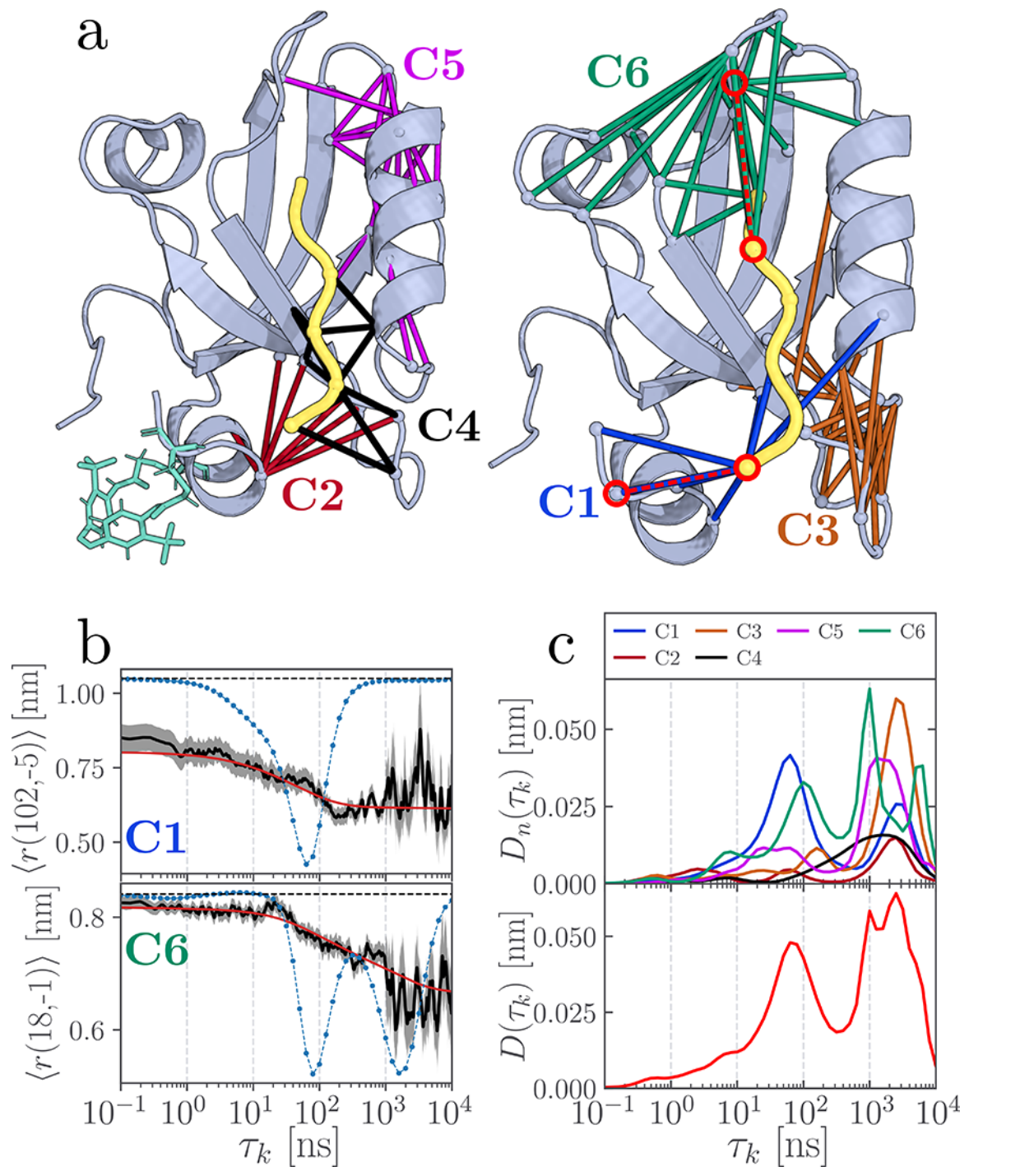

**Figure 3.** Contact clusters (a) and nonequilibrium response (b, c) of a photoswitchable PDZ3 domain (PDZ3L6) with a ligand that is one residue longer than the PDZ3 shown in Figure 1. See the caption of Figure 1 for additional information.

## PDZ2S: Cross-linking the Binding Pocket

Beyond the PDZ3 domains—for which recurring patterns in the contact clusters were seen—we also studied two PDZ2 variants for which the $\alpha_3$-helix is absent. In the first one, termed PDZ2S ("S" for switch), the photoswitch links residue Azo22 at the $\beta_2$-sheet with Azo77 in the $\alpha_2$-helix. Upon *cis*-to-*trans* photoisomerization, the photoswitch therefore expands the (empty) binding pocket from an unbound-like state to a bound-like conformation, which represents a rather strong and direct perturbation. Providing a simple model of ligand binding, PDZ2S was studied by time-resolved infrared spectroscopy[29] and accompanying MD simulations.[15]

Using the nonequilibrium simulations of Buchenberg et al.,[15] we identified 330 contact distances, for which a MoSAIC analysis was carried out (Table S7). Figure 4a shows the resulting five contact clusters, which overall resemble those of PDZ3 (Figure 1a), but differ in various aspects. Notably, the functionally important $\alpha_3$-helix of PDZ3 is not present in PDZ2 domains, hence clusters C1 and C2—both connected to the $\alpha_3$-helix—are not to be found in PDZ2. While both systems otherwise share the same secondary structures (2 $\alpha$-helices, 6 $\beta$-strands), their sequence similarity is only 36% (Table S1), such that an exact correspondence of clusters is neither expected nor of particular relevance. Nevertheless, we identify similar clusters in both systems: C3 at the $\beta_2\beta_3$-loop, C6 around the $\beta_1\beta_2$-loop, and C7 at the $\alpha_1$-helix, although the latter contains significantly more contacts in PDZ2S. Especially the region around C6 and C7 has shown to be very important in allosteric interactions.[53,54] Responding directly to the photoswitching of the binding pocket, we further note the presence of a large cluster C4 at the $\alpha_2$-helix as well as a small additional cluster C8 involving the $\beta_2$ and $\beta_3$ strands.

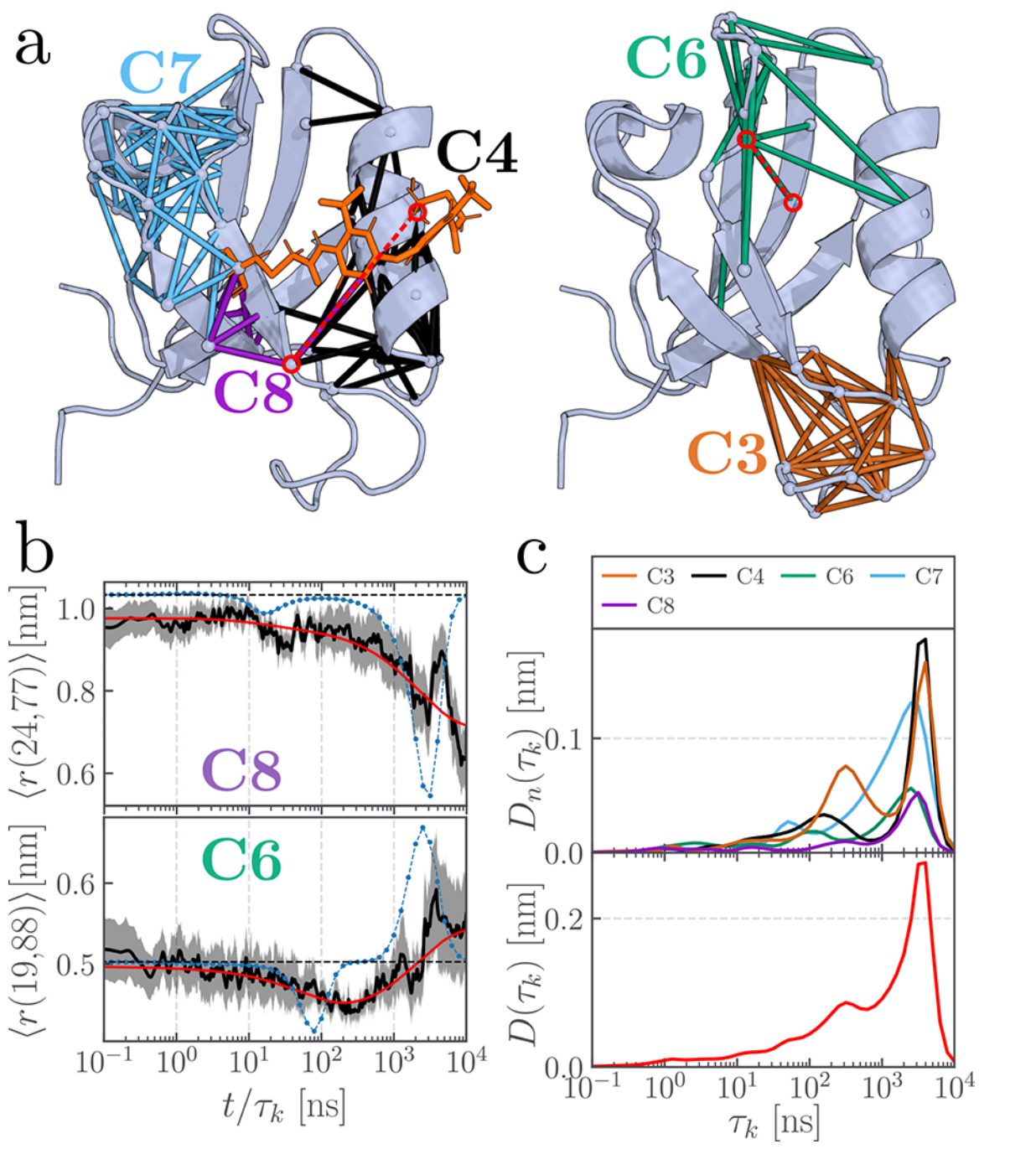

**Figure 4.** Contact clusters (a) and nonequilibrium response (b, c) of a PDZ2 domain (PDZ2S) featuring a photoswitch across its binding pocket. See the caption of Figure 1 for additional information.

To explore the nonequilibrium response of PDZ2S following photoswitching of its binding pocket, we again analyze the time scales of all contact distances and compute the dynamical content across all clusters (see Figure 4b,c). The overall dynamical profile reveals two dominant peaks around 300 ns and 4 $\mu$s as well as weak features at 1 and 10 ns, which are consistent with experimental observations.[29] The 10 ns feature corresponds to the photoinduced opening of the binding pocket, as indicated by a first time scale in the contact distance $r(24, 77)$ within cluster C8 (Figure 4b). The 10 ns peak appears relatively small, as only a few contacts span the binding pocket region, see Table S7. The same is true for the 1 ns time scale, which is caused by several contacts in clusters C4, C6 and C8 (Figure S11), which are in direct proximity of the anchors of the photoswitch. Between 200 and 400 ns, we observe a broad peak that reflects the rearrangement of clusters C3 and C4 in response to the opening of the binding pocket. The subsequent structural relaxation of these clusters gives rise to the prominent peak at 4 $\mu$s, representing the fully opened binding pocket.

Interestingly, we also detect response of the two distant clusters C6 and C7, see Figures 4b and S11. For example, distance $r(19, 88)$ of C6 shows a transient contact change involving the $\beta_1\beta_2$-loop, while several distances of C7 report on the ongoing destabilization of the short $\alpha_1$-helix. These contacts produce small peaks at 50 and 150 ns, respectively, and contribute additional features at 2 $\mu$s, thereby adding a shoulder to the dominant 4 $\mu$s peak in the overall dynamical content. Thus, despite the considerable perturbation of PDZ2S by the cross-linked photoswitch, the overall structure of the contact clusters and their dynamic response in many aspects resemble the behavior previously observed for photoswitched PDZ3. In particular, we again identify long-range coupling between the initially excited clusters (C4 and C8) and the distant clusters (C3, C6 and C7) at opposite ends of the protein.

## PDZ2L: Photoswitching the Ligand

The second PDZ2 variant we study contains a relatively long ligand of 16 residues, where only about half of the residues are bound to the binding pocket, see Figure 5a. Termed PDZ2L ("L" for ligand), it differs from the other systems in that the photoswitch is attached to the ligand (at residues Azo(−1) and Azo(−6)) rather than to the protein itself.[30] Photoswitching azobenzene from *trans*-to-*cis* results in a squeezing of the ligand, which destabilizes its contacts in the binding pocket and ultimately leads to its unbinding. Conversely, *cis*-to-*trans* photoswitching facilitates the binding of the ligand to PDZ2L.[30] Compared to PDZ2S where the binding of a ligand was mimicked by the photoinduced opening of the binding pocket, the ligand-induced binding and unbinding represents a smaller and more realistic perturbation of the system.

Utilizing 441 contact distances in the MoSAIC analysis, we find six functional clusters as presented in Figure 5a. While cluster C3, C5, C6, and C7 reappear similar to the other PDZ domains, we in particular obtain clusters C4 and C6', which are located in the binding-pocket region and link ligand residues Val(0) to Glu(−7) to the PDZ2 domain. (While C6' is closely related to C6, it emerges as separate cluster in the MoSAIC analysis, see Table S8.) This is a direct result of the ligand unbinding being the largest and most crucial process which occurs in the system.

Monitoring the initial destabilization of the photoswitched ligand, clusters C4 and C6' exhibit an early response, yielding a broad dynamical content with peaks at 8, 30, and 150 ns of increasing amplitude (Figure 5c). As an example monitoring this process, Figure 5b shows the evolution of contact distances $r(72, -5)$ of cluster C4, while distance $r(76, -6)$ reveals the time-delayed response of cluster C6'. The dominant peak of the dynamical content at 4 $\mu$s reflects the rearrangement of virtually all contact clusters, which indicates that the ligand has left its initial binding pose. Although hardly visible in the dynamical content, we again find several contact distances that respond already on a 1 ns time scale, see Figure S13.

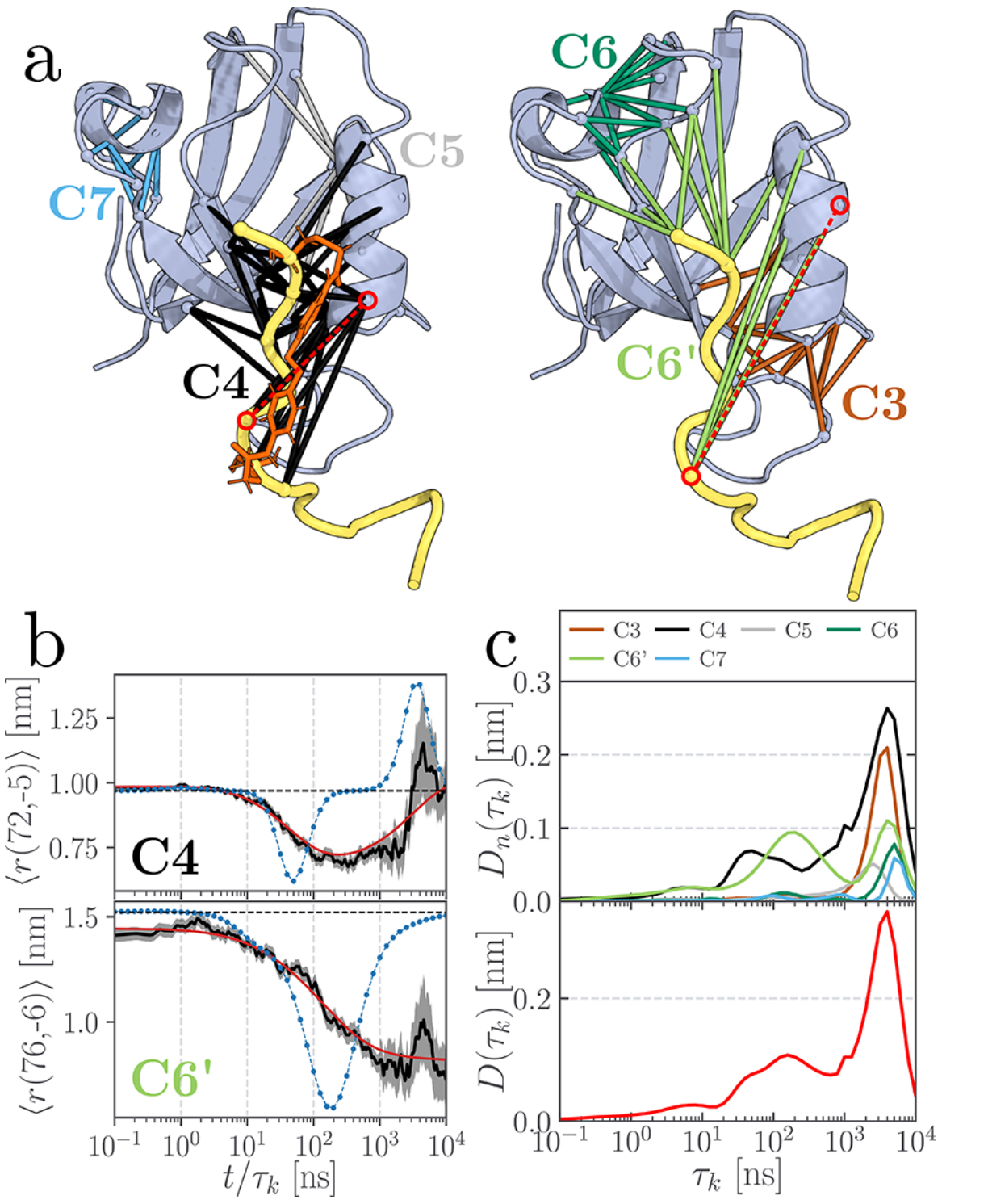

**Figure 5.** Contact clusters (a) and nonequilibrium response (b,c) of a PDZ2 domain (PDZ2L) featuring a photoswitched ligand across its binding pocket. See the caption of Figure 1 for additional information.

As mentioned above, Bozovic et al.[30] also studied the reverse reaction, i.e., the binding of the ligand to PDZ2L following *cis*-to-*trans* photoswitching. To simulate this process, the rather heterogeneous conformational distribution of the *cis* equilibrium state was sampled from the last microseconds of the *trans*-to-*cis* nonequilibrium simulations. Taking randomly chosen snapshots from this distribution, *cis*-to-*trans* nonequilibrium simulations were performed. Given that initially the unbound ligand can be anywhere around the protein, and that the length of the trajectories ($100 \times 1$ $\mu$s, $10 \times 10$ $\mu$s) is on average too short to ensure the complete binding of the ligand, the overall sampling quality of the photoinduced ligand binding is clearly inferior to the one of the above-discussed unbinding process. Nonetheless, the data allow for some general qualitative observations, which are presented in the SI (Figure S14 and Table S9) for completeness. Using 403 contacts, the MoSAIC analysis resulted in eight major clusters, where C3 and C6 are again similar as found before. As may be expected, there are several prominent clusters spanning over the binding pocket, in particular C1, C2 and C4 which connect the ligand to the protein. In line, we find that these clusters as well as C6 exhibit an early response at 1 and 60 ns, before all clusters peak at ∼3 $\mu$s. Although the approach of the ligand to the binding pocket facilitates many ligand-protein contacts, the binding process can be still described via a set of contact clusters.

## CONCLUSIONS

We have shown that the dynamic decomposition of proteins into MoSAIC clusters reveals a dynamics-based modular architecture that enables long-range communication. To assess the validity and generality of this concept, we have adopted four photoswitchable PDZ domains and investigated how different domains, ligands and perturbations affect both the clusters and their dynamical evolution.

MoSAIC analyses across these systems have revealed several recurring contact clusters that represent shared flexible structural modules. Notably, cluster C3 is located at the $\beta_2\beta_3$-loop, C6 around the $\beta_1\beta_2$-loop as well as C7 at the $\alpha_1$-helix. Additional clusters emerge in response to the photoswitch perturbation—specifically, C1 and C2 (at $\alpha_3$) in PDZ3 domains and C4 (at $\alpha_2$) in PDZ2 domains. Although the precise contacts comprising a given cluster may vary between systems, exact correspondence is neither expected nor essential. We note that residues that reiteratively occur in one or several contact clusters are key to the allosteric communication.[18,19,55] Mutations of these residues will change the allosteric coupling, as was demonstrated for the modification of the ligand in PDZ3 (Figures 1 and 3). Furthermore, equilibrium MD simulations of PDZ3WT confirm that contact clusters can be identified under equilibrium conditions, reinforcing that such clusters are intrinsic to the protein's architecture.

To investigate how contact clusters mediate protein dynamics and function, we have analyzed the nonequilibrium response of all systems by computing the dynamical content [eq 2], where the peak positions indicate characteristic time scales of the protein's response and can be linked to the motions of individual contact clusters. By comparing two PDZ3 domains with slightly different ligands (Figures 1 and 3) and two PDZ2 domains—one containing a cross-linked photoswitch (PDZ2S, Figure 4) and one with a photoswitched ligand (PDZ2L, Figure 5)—we identified several common features in their dynamical content. Overall, at least three distinct time scales emerge:

- 1−10 ns, corresponding to the local stress initially induced by the photoswitch,
- 100−300 ns, reflecting the response of more distant residues,
- a few $\mu$s, reporting on a global conformational response of the protein.

While a local short-time response to photoswitching is generally expected, it became most evident in the dynamical content of PDZ3, where the initially perturbed cluster C1 contains several contacts in close proximity of the photoswitch. Most interestingly, across all systems, a coupling between the perturbed end of the PDZ domain (e.g., clusters C1 or C4) and more distal regions (e.g., clusters C6 and C7) occurs on a time scale of about 50−200 ns. In PDZ3, this long-range communication has been shown to be mediated by the ligand, which forms multiple contacts linking these clusters.[46] In general, this coupling likely involves several competing mechanical pathways mediated by rigid secondary structures. For the isolated PDZ domains studied here, this long-range interaction does not appear to serve a specific functional purpose. However, if the $\beta_1\beta_2$-loop or $\beta_3\alpha_1$-loop of clusters C6 and C7 were to interact with a nearby molecule (such as another protein or cofactor, as discussed in ref 22), the formation of new contacts could stabilize the allosteric response of cluster C6. This would lead to an allosteric transition in C6, triggered by C1. For example, the Rho GTPase Cdc42 has been shown to allosterically regulate the Par-6 PDZ domain by binding to $\beta_1$ and $\alpha_1$.[53]

The consistent presence of a dominant microsecond-scale response involving nearly all contact clusters is remarkable. While the sub-$\mu$s dynamics discussed above is specific to the system, virtually all contact distances across all systems exhibit this long-time behavior, see Figures S2, S9, S11 and S13 in the Supporting Information. Supporting the view that conformational transitions in proteins are largely governed by diffusive processes,[56−60] the microsecond dynamics of these distances can be well described by a power-law dependence, $r(t) \propto t^\alpha$. This is demonstrated for a number of examples in Figure S17, which reveals that the exponent $\alpha$ typically ranges between 0.3 (anomalous or subdiffusive motion) and 0.5 (normal diffusion). Since our simulation trajectories are limited to 10 $\mu$s, the observed microsecond-scale response of nearly all distances may indicate that some distances have not yet equilibrated. [Generally, when extending MD simulations (say, from 1 to 10 $\mu$s) in order to account for a slow (say, 2 $\mu$s) process, the largest time constant predicted by the time scale analysis will increase from a too short value to the correct result, because the short MD data sees only the initial phase of this process.] This observation aligns with experimental findings,[28−30] which suggest that the overall photoinduced conformational transition in PDZ domains is not completed before 100 $\mu$s.

In this Perspective, we have focused on photoswitchable PDZ domains, whose nanosecond−to−microsecond response times enable direct observation of the allosteric transition. Ongoing work extends this analysis to larger allosteric systems such as the tetracycline repressor TetR and the heat shock protein Hsp90, for which contact clusters can be readily derived from available microsecond-scale equilibrium simulations.[61,62] Identifying which contacts change during a conformational transition also enables the construction of mechanism-informed biasing coordinates for enhanced sampling techniques.[63,64] In this way, physics-based, mechanism-guided sampling strategies may effectively complement data-driven machine learning approaches for exploring allosteric mechanisms in proteins.

## ■ METHODS

### MD Simulations

All simulations used the GROMACS v2020 software package,[65] the Amber99SB*ILDN force field[66−68] and the TIP3P water model,[69] and were run at a temperature of 300 K, with a coordinate write out time of 20 ps. Table 1 list all systems and simulations, while specific MD details for PDZ3, PDZ2S, and PDZ2L are given in refs 31,15, and 30, respectively. For all systems, first a set of $\mu$s-long equilibrium simulations were conducted, from which statistically independent starting configurations for the nonequilibrium runs were obtained. The latter used the potential energy surface switching method by Nguyen et al.,[70] to mimic the initial photoisomerization of the azobenzene photoswitch. Further details are given for each specific system in the SI.

### MoSAIC Clustering

In a first step, we determined the inter-residue contacts of each system. Excluding the two nearest neighbors ($|i - j| > 2$ for residues $i$ and $j$), this is achieved by requesting that the shortest heavy-atom distance between the two residues is below 0.45 nm for at least 10% of the total simulation time.[42,43] (The validity of our distance threshold of 0.45 nm was recently confirmed in a detailed study by Yao et al.[44] The value of contact population threshold of 10% is not critical, but mainly helps to exclude irrelevant contacts from the analysis.) To distinguish collective motions underlying functional dynamics and uncorrelated motions, MoSAIC[45] ("Molecular Systems Automated Identification of Cooperativity") analysis calculates the linear or nonlinear[71] correlation matrix between all contact distances. Dealing with scalar variables (e.g., distances), our recent studies[45,71] suggest that it is sufficient to restrict ourselves to the linear Pearson correlation, which is not only much easier to compute but generally also easier to interpret. Only for a highly perturbed systems such as PDZ2S, normalized mutual information seemed to be advantageous (Figure S10). To rearrange this matrix in an approximately block-diagonal form, we employ a community detection technique called Leiden clustering.[72] Using the constant Potts model, the user-chosen resolution parameter $\gamma$ ($0 < \gamma \leq 1$) determines the average correlation of the resulting blocks or clusters: larger values of $\gamma$ yield many highly correlated clusters, while smaller values give few but larger and less correlated clusters. Since MoSAIC is very robust with respect to sparse data, it is sufficient to use only every fifth frame of a trajectory.

### Time Scale Analysis

Each averaged distance is fitted by eq 1 using a maximum-entropy method[48] that minimizes $\chi^2 - \lambda S_{ent}$, where $\chi^2$ is the usual root-mean-square deviation and $S_{ent}$ the entropy-based regularization factor. The regularization parameter $\lambda$ needs to be chosen prior to the analysis and controls over- and under-fitting. Means to determine $\lambda$ and check for the statistical robustness of the associated multiexponential fits were recently discussed in refs 60 and 73. For each system, one common value is chosen: $\lambda^{PDZ3} = 50$, $\lambda^{PDZ3L6} = 100$, $\lambda^{PDZ2L} = 200$ and $\lambda^{PDZ2S} = 50$. We improve the convergence, precision and stability of the fit by several measures: (1) A linear interpolation between all neighboring frames is employed increasing the number of frames by a factor of 4. (2) The frames are transformed to be log-spaced along the time-axis which ensures that each decade contributes equally with about 750 frames. (3) To suppress fast fluctuations of the data, a low-pass Gaussian filtering[52] is applied with a standard deviation of $\sigma = 6$ frames. (4) To improve the fit at the upper boundary, the data is extended by one additional order of magnitude by a constant value, derived as average over half of the previous decade.[73]

## ■ ASSOCIATED CONTENT

### Data Availability Statement

The clustering package MoSAIC is available at our group homepage https://www.moldyn.uni-freiburg.de/software.html. Trajectories and simulation results are available from the authors upon reasonable request.

### Supporting Information

The Supporting Information is available free of charge at https://pubs.acs.org/doi/10.1021/acs.jpcb.5c07737.

Sequence and secondary structures of all considered proteins, the MoSAIC correlation matrix, the list of contacts included in the clusters, and their nonequilibrium time evolution (PDF)

## ■ AUTHOR INFORMATION

### Corresponding Author

**Gerhard Stock** − *Biomolecular Dynamics, Institute of Physics, University of Freiburg, 79104 Freiburg, Germany;* orcid.org/0000-0002-3302-3044; Email: stock@physik.uni-freiburg.de

### Authors

**Emanuel Dorbath** − *Biomolecular Dynamics, Institute of Physics, University of Freiburg, 79104 Freiburg, Germany;* orcid.org/0009-0003-0148-0907

**Fabian Rudolf** − *Biomolecular Dynamics, Institute of Physics, University of Freiburg, 79104 Freiburg, Germany*

**Adnan Gulzar** − *Biomolecular Dynamics, Institute of Physics, University of Freiburg, 79104 Freiburg, Germany*

Complete contact information is available at: https://pubs.acs.org/10.1021/acs.jpcb.5c07737

### Notes

The authors declare no competing financial interest.

### Biographies

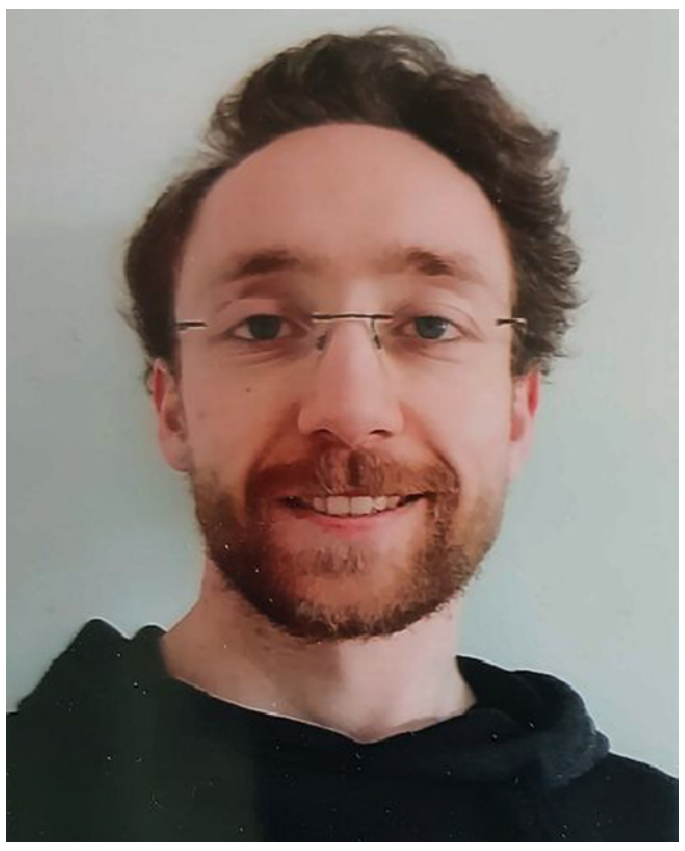

Emanuel Dorbath is a postdoctoral researcher at the University of Freiburg, Germany. He received his B.Sc., M.Sc., and Ph.D. in Physics from the University of Freiburg. His research is centered on the study of hierarchical dynamics and allosteric communication in proteins, with a particular focus on nonequilibrium processes.

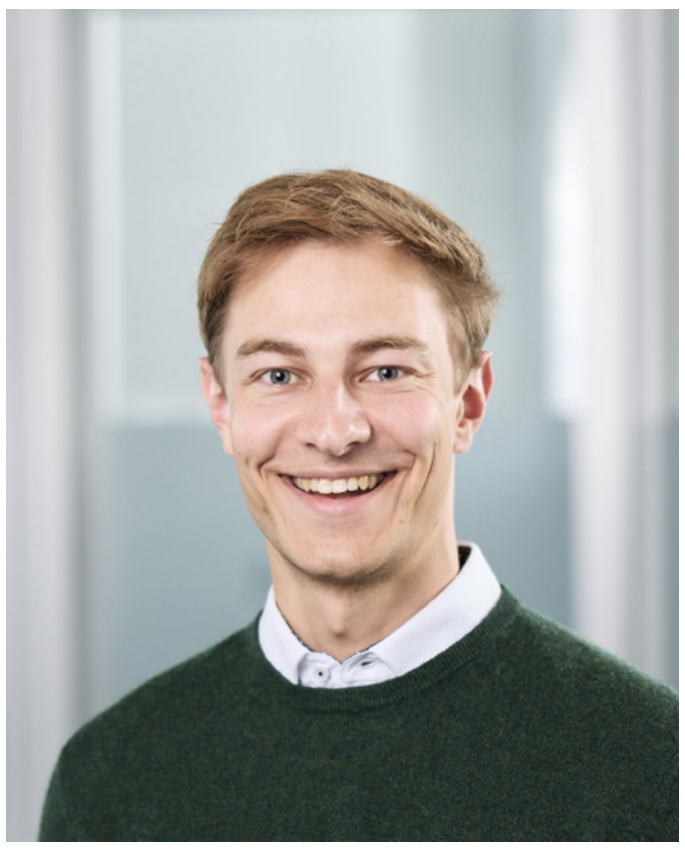

Fabian Rudolf received his B.Sc. in Physics from Kiel University and his M.Sc. in Applied Physics from the University of Freiburg. His research focuses on allosteric communication in proteins.

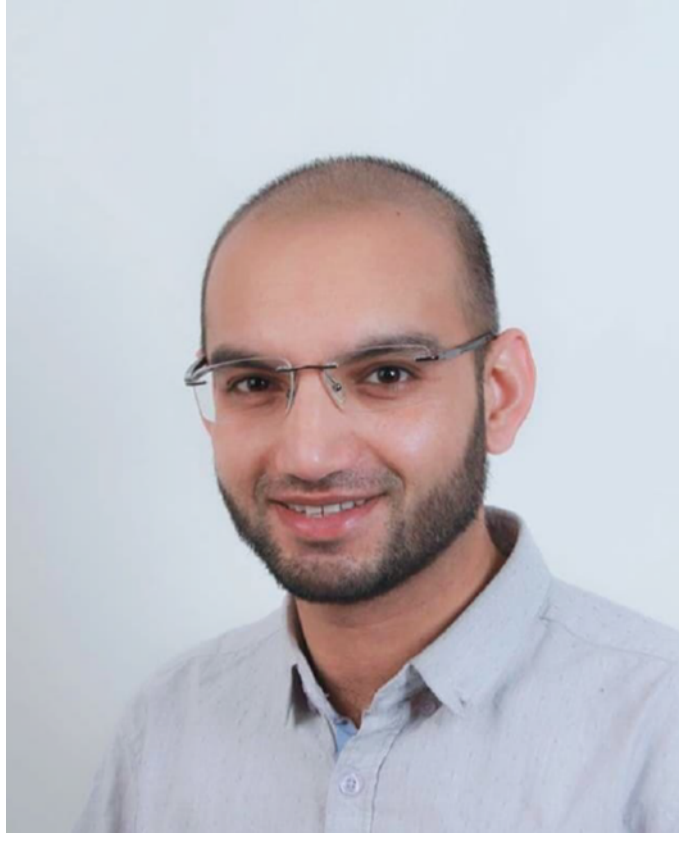

Adnan Gulzar is a senior scientist at Roche in Penzberg, Germany. He earned his Ph.D. in Chemistry from the University of Freiburg, where his research focused on allosteric communication and energy transport in proteins. His work centers on computational modeling of biomolecular dynamics, with particular interests in allosteric regulation, enzymatic reaction mechanisms, and theoretical spectroscopy.

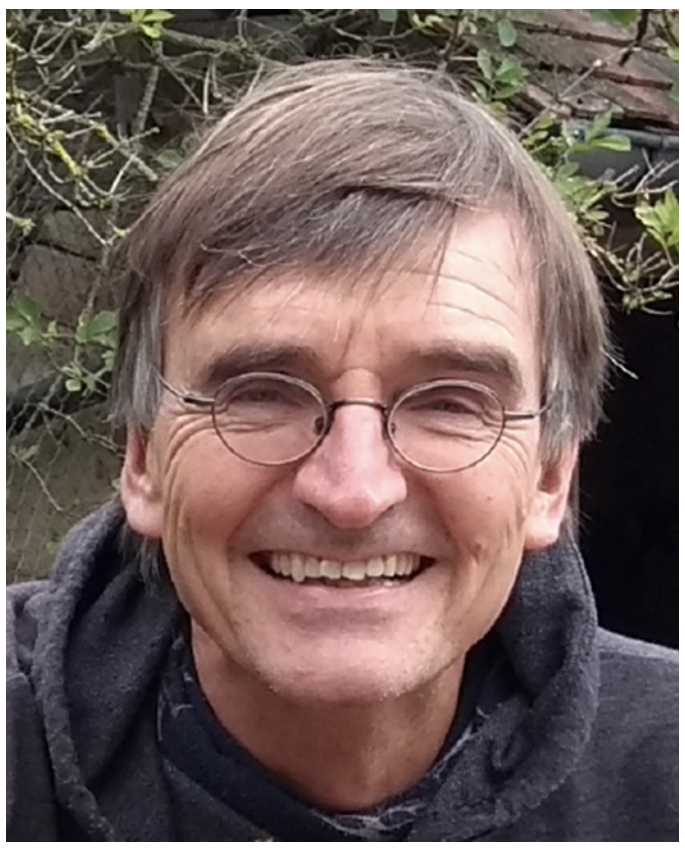

Gerhard Stock is a Professor of Physics at the University of Freiburg, Germany. He earned a Diploma in Physics in 1987 and a Ph.D. in Theoretical Chemistry in 1990 from the Technical University of Munich, where he worked with Wolfgang Domcke. He then carried out postdoctoral research with William H. Miller at the University of California, Berkeley, before returning to the Technical University of Munich to complete his Habilitation in 1996. After holding a Heisenberg Professorship at the University of Freiburg, he accepted in 2000 the Chair of Theoretical Chemistry at the University of Frankfurt and later moved back to Freiburg in 2009. His research focuses on the theory and simulation of elementary biomolecular processes and their time-resolved detection, with particular emphasis on developing new multiscale simulation methods and strategies to reduce the complexity of nonequilibrium phenomena.

## ■ ACKNOWLEDGMENTS

The authors thank Peter Hamm, Steffen Wolf, Georg Diez, Ahmed Ali, Camilla Sordi, and Sofia Sartore for helpful comments and discussions. This work has been supported by the Deutsche Forschungsgemeinschaft (DFG) within the framework of the Research Unit FOR 5099 "Reducing complexity of nonequilibrium"(project No. 431945604), the High Performance and Cloud Computing Group at the Zentrum für Datenverarbeitung of the University of Tübingen and the Rechenzentrum of the University of Freiburg, the state of Baden-Württemberg through bwHPC and the DFG through grant no INST 37/935-1 FUGG (RV bw16I016), and the Black Forest Grid Initiative.

## ■ REFERENCES

## NOTE ADDED AFTER ASAP PUBLICATION